\documentclass[twocolumn,twoside]{revtex4}
\usepackage{graphicx}
\usepackage{fancyhdr}
\begin{document}

\title{Overview of R(D) and R(D*)}

%

\author{R.~Cheaib\\
on behalf of the Belle II experiment}
\affiliation{DESY, Hamburg, Germany }

\begin{abstract}
In this talk, an overview of the latest $R(D^{(*)})$ measurements performed at the Belle and  LHCb experiments is presented. The main approach and methodology of each measurement is discussed, along with the current limitations and potential improvements. Furthermore, the prospects for $R(D^{(*)})$ measurements at Belle II and the tools developed to improve their overall precision are also presented. 
\end{abstract}

\maketitle

\thispagestyle{fancy}


\section{Introduction}
\begin{figure}[h]
\centering
\includegraphics[width=80mm]{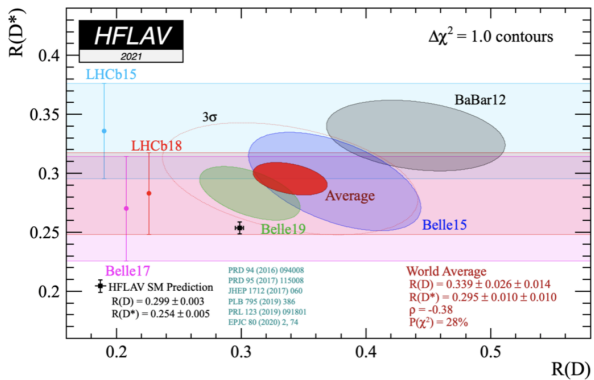}
\caption{Current status of $R(D^{(*)})$ measurements and the average as given by the Heavy Flavour Averaging group \cite{HFLAV}.} 
\label{HFlav}
\end{figure}

Lepton flavour universality (LFU) is a fundamental symmetry of the Standard Model (SM). It implies that gauge interactions of the three generations of leptons: $e, \mu, \tau$,  are identical once the mass difference is accounted for. Any violation of LFU is thus a smoking gun signal of new physics .
To this effect, semileptonic $B$ decays are an invaluable portal for LFU tests. 
$R(D^{(*)}$ is  defined by the following equation:
\begin{equation}
    R(D^{(*)}=\frac{\mathcal{B}(B\rightarrow D^{(*)}\tau\nu_{\tau})}{\mathcal{B}(B\rightarrow D^{(*)}\ell\nu_{\ell})}
\end{equation}
where here $\ell=e,\mu$. The numerator is often referred to as the signal mode, while the denominator is referred to as the normalization mode. The ratio allows for many uncertainties to cancel and  is determined to high precision in the SM. This ratio has been measured by BaBar\cite{babar_RDstar}, Belle \cite{belle_hadtag_leptau_RDstar} \cite{belle_hadtag_hadtau_RDstar} \cite{belle_sl_RDstar} and LHCb\cite{lhcb_RDstar_muonic} \cite{lhcb_RDstar_hadronic} and a combined deviation of 3.1$\sigma$ deviation, shown in Fig \ref{HFlav}, from the SM expectation has been observed. This has sparked the interest of the particle physics community and various new physics models have been developed as potential explanations to the observed deviation. 

\section{$R(D^*)$ measurements at LHCb}
At the LHCb experiment, $R(D^*)$ has been measured with leptonic and hadronic $\tau$ decay modes. Because of the neutrino in its final state, the $\tau$ momentum four vector cannot be exactly determined. Instead, vertexing information obtained from the LHCb detector \cite{LHCb_detector} is used to approximate $B$ kinematics and suppress leading backgrounds. For the leptonic measurement,  only the $\tau^+\rightarrow \mu^+ \nu_{\mu}\bar{\nu}_{\tau}$ is accessible.

\subsection{$R(D^*)$ with $\tau^+\rightarrow \mu^+ \nu_{\mu} \bar{\nu}_{\tau}$ \cite{lhcb_RDstar_muonic}}
\begin{figure*}[t]
\centering
\includegraphics[width=140mm]{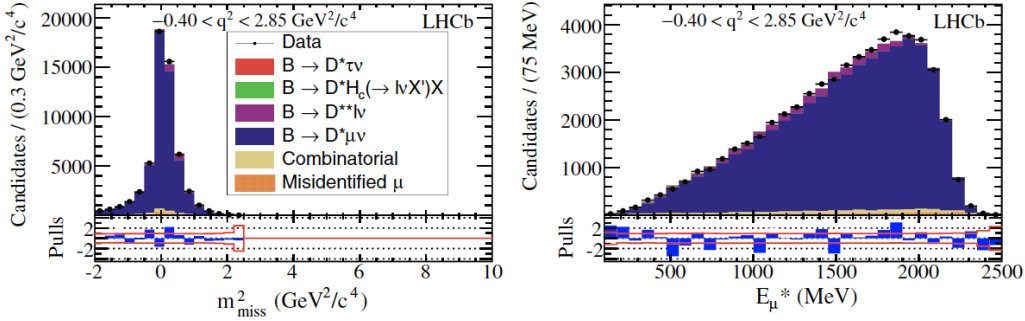}
\caption{Projections of the fit to $E_{\mu}^*$ and $m^2_{miss}$ of LHCb data overlaid with the simulated signal and background MC distributions.} \label{Fit_RDstar_muonic_lhcb}
\end{figure*}

This measurement performed with 3.0 fb$^{-1}$ of LHCb data collected during 2011-2012. To determine $R(D^*)$, a common reconstruction procedure is emplored for both the signal mode $B^0\rightarrow D^{*-}\tau^+\nu_{\tau}$  and the normalization mode $B^0\rightarrow D^{*-}\mu^+\nu_{\mu}$, where $D^{*+}\rightarrow D^0 \pi^+, D^0 \rightarrow K^-\pi^+$. A multivariate (MVA) algorithm, based on the track separation from the primary vertex (PV), the track angle, etc...,  is developed to distinguish whether a charged track originated from the signal $B$, $B_{sig}$ or the rest of event (ROE). To seperate between signal and normalization events, variables such as $E_{\mu}^*$: the energy of the muon in the center-of-mass (CM frame), $q^2$: the square of the momentum imparted to the lepton and its neutrino given by $(p_{B}^2-p_{D^*}^2)$, and $m^2_{miss}$: the missing mass squared given by $(p_{B}^2-p_{D^*}^2-p_{\mu}^2)$. Here, $p_{B}$,$p_{D^*}$, and $p_{\mu}$ are the four momentum vectors of the $B_{sig}$, $D^*$, and $\mu$ respectively. 
\begin{figure}[h]
\centering
\includegraphics[width=80mm]{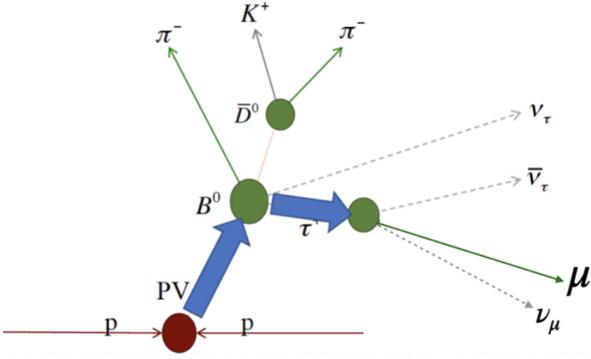}
\caption{Depiction of a $\bar{B}^0\rightarrow D^{*+}\mu^-\nu_{\mu}$ event at LHCb and how the $B_{sig}$ rest frame is determined.} \label{LHCBmuonic+depiction}
\end{figure}
To determine $q^2$ and $m^2_{miss}$, the rest frame of the $B_{sig}$ is required. This is done by determining the direction of $B_{sig}$ and its corresponding momentum along the beamline. As depicted in Fig. \ref{LHCBmuonic+depiction}, the direction of  the $B_{sig}$ is determined from the unit vector between the PV and the $B$ decay vertex. In addition, the component of the $B$ momentum along the beam axis, $(p_B)_z$ is approximated using the relation $(p_{B})_z=\frac{m_B}{m_{reco}}(p_{reco})_z$, where $m_B$ is the mass of the $B_{sig}$ and $m_{reco}$ and $(p_{reco})_z$ are the mass and  z-component of the momentum of the $D^*\mu$ system respectively. After all signal selection cuts, leading backgrounds from 
semileptonic decays  to excited charm states $B\rightarrow D^{**}\ell\nu$ , double charm $B$ decays $B\rightarrow D^{(*)}H_c, H_c\rightarrow \mu\nu X$, and B decays with hadrons misidentified muons are still present.
A maximum likelihood fit is performed  to $E_{\mu}^*$ and $m^2_{miss}$  in 4 bins of $q^2$ to extract the signal, normalization and background yields, as shown in Fig \ref{Fit_RDstar_muonic_lhcb}. The resulting value is $R(D^*)=0.336\pm0.027 \rm(stat) \pm0.030 \rm(sys)$, where the leading systematic error is from the limited size of the Monte Carlo samples as well as from the template shapes of background events in the final fit. The result is 1.7$\sigma$ over the SM expectation. This is the first  measurement of $R(D^*)$ at a hadronic collider. Improved modeling of background events, using dedicated control samples, can further decrease the overall systematic uncertainty. Future simultaneous measurement of $R(D)$ and $R(D^*)$ at LHCb with Run 1 data and Run 2 data are in progress, where the latter is 4 times the available statistics. 

\subsection{$R(D^*)$ with $\tau^+\rightarrow \pi^+ \pi^-\pi^{+} (\pi^0) \bar{\nu}_{\tau}$\cite{lhcb_RDstar_hadronic}}
\begin{figure}[h]
\centering
\includegraphics[width=80mm]{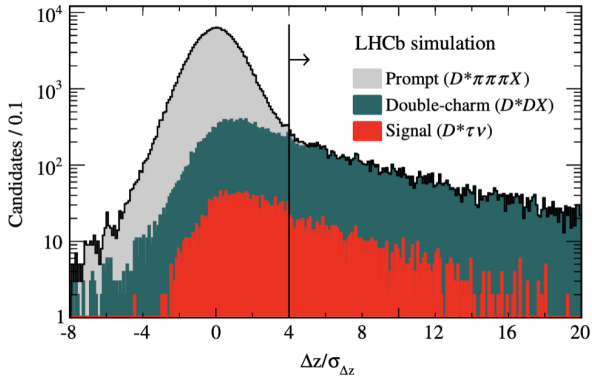}
\caption{The distribution of $\Delta z/\sigma_z$ for simulated background and signal events. A cut at 4 significantly reduces the abundance of $B \rightarrow D^* 3 \pi X$ and $B\rightarrow D D^* (X)$.  } \label{deltaz_LHCB_hadronic}
\end{figure}

\begin{figure*}[t]
\centering
\includegraphics[width=140mm]{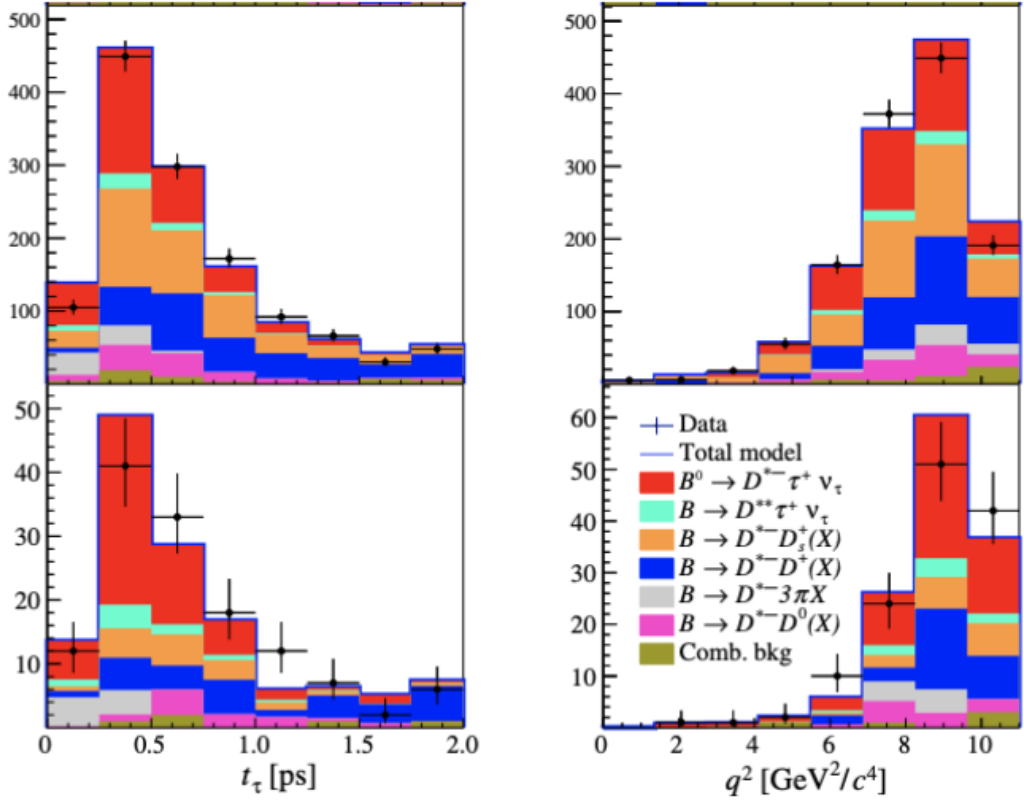}
\caption{Projections of the fit to the tau decay time, $t_{\tau}$ and $q^2$ in two bins of the BDT output. The BDT output increases from top to bottom, along with the abundance of signal events. } \label{tau_hadronic_lhcb_BDT}
\end{figure*}

To measure $R(D^*)$ with 3 prong $\tau$ decays, the normalization mode is chosen to be $B^0 \rightarrow D^{*-} 3\pi$ and the ratio $\kappa(D^*)=\frac{B^0\rightarrow D^{*-}\tau^+\bar{\nu}_{\tau}}{B^0 \rightarrow D^{*-} 3\pi}$ is first determined. Then, $R(D^*)$ can be measured with the following:
\begin{equation}
    R(D^*)=\kappa(D^*) \times \frac{\mathcal{B}(B^0\rightarrow D^{*-}3\pi)}{\mathcal{B}(B^0\rightarrow D^{*-}\mu^+\nu_{\mu})}
\end{equation} 
The challenge when measuring $R(D^*)$ with hadronic $\tau$ decays is the large background originating from $B \rightarrow D^* 3 \pi X$ and $B\rightarrow D D^* (X)$. This background is almost 100 times the size of the signal and can be suppressed with  a requirement on $\Delta z/\sigma_z$,shown in Fig \ref{deltaz_LHCB_hadronic}, where $\Delta z$ is the distance along the beam line from the $3\pi$ vertex to that of the $B$ and $\sigma_z$ is the corresponding uncertainty. 
Furthermore, double charm backgrounds, including $D_s$ decays, are further suppressed using a Boosted Decision Tree (BDT) algorithm. The BDT uses variables such as the resonant structures of the $\tau$ decay, the kinematic properties of the $B$ decay  etc ... to separate between signal and background events. 
The final  signal yield is then extracted using a fit to the $\tau$ decay time, $t_{\tau}$, and $q^2$, in 4 bins of the BDT output as shown in Fig \ref{tau_hadronic_lhcb_BDT}.  The resulting value  is $R(D^*)=0.291\pm0.019 \rm(stat) \pm0.026 \rm(sys) \pm 0.013 \rm(ext)$, where the third uncertainty is due to the limited knowledge of the external branching fractions.  The limited knowledge of the $D_s^+$ decay model is also one of the leading systematic uncertainties for this measurement. External measurements of the double charm decays can improve the precision of this result.A subsequent measurement of $R(D^*)$ with hadronic $\tau$ decays  is planned at LHCb with Run 2 data, along with a measurement of the $D^*$ polarization in $B^0\rightarrow D^{*-}\tau^+ \nu_{\tau}$. In addition, a recent measurement of $R(\Lambda_c)=\frac{\mathcal{B}(\Lambda_b \rightarrow \Lambda_c \tau^+ \nu_{\tau})}{\mathcal{B}(\Lambda_b \rightarrow \Lambda_c \mu^+ \nu_{\mu})}$ has recently performed, where the decay $\Lambda_b \rightarrow \Lambda_c \tau^+ \nu_{\tau}$ is observed for the first time with 6.1 $\sigma$ significance. The result is consistent with the SM and constrains new physics models that predict high values of $R(\Lambda_c)$\cite{lhcb_Rlamdac}.

\section{$R(D^{(*)})$ measurements at Belle}
At $e^+e^-$ colliders, $B$ tagging algorithms are employed to allow for precision  $R(D^{(*)})$ measurements.  With $B$ tagging, one of the $B$ mesons produced in $\Upsilon(4S)\rightarrow B\bar{B}$ decay, referred to as $B_{\rm tag}$, is exclusively reconstructed using hadronic or semileptonic modes. The remaining information in the event is attributed to the other $B$, $B_{\rm sig}$, on which the search for $B\rightarrow D^{(*)}\tau\nu_{\tau}$ is performed. This approach determines the four momentum of the $B\bar{B}$ pair and thus allows access to the full event kinematics, which is  particularly useful for decays with neutrinos. At Belle, $R(D^{(*)})$ has been measured with hadronic and semileptonic $B$-tagging, with hadronic and leptonic $\tau$ decays with using full Belle dataset, which corresponds to 711 fb$^{-1}$.  
\subsection{$R(D^{(*)})$ with  hadronic tagging and leptonic $\tau$ decays\cite{belle_hadtag_leptau_RDstar}}
\begin{figure}[h]
\centering
\includegraphics[width=80mm]{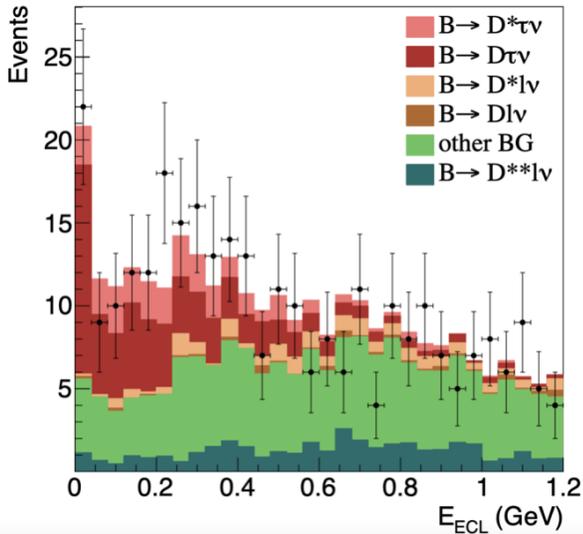}
\caption{The distribution of $E_{ECL}$ for simulated  signal and background, events. The Belle data is shown as points with error bars.  } \label{Eecl_belle}
\end{figure}

\begin{figure*}[t]
\centering
\includegraphics[width=140mm]{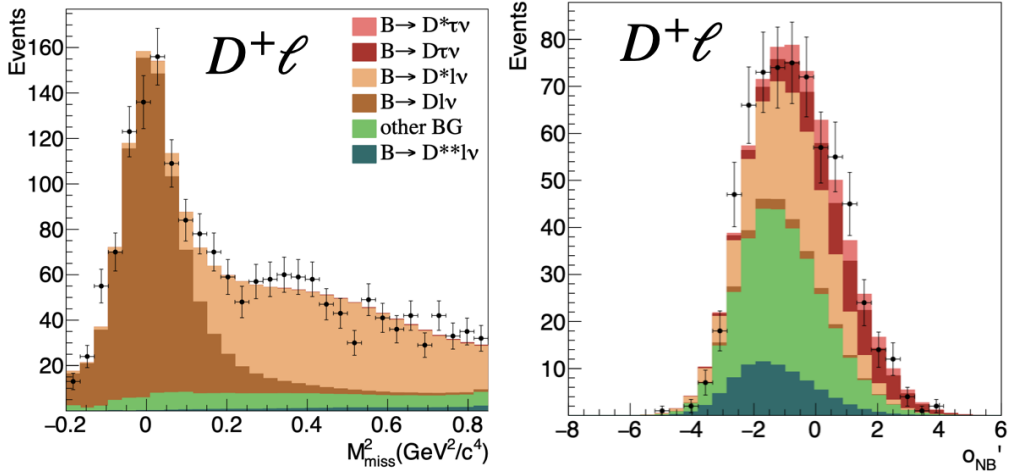}
\caption{For the $B^0\rightarrow D^-\ell^+\nu_{\ell}$ channel, projections of the fit to $m^2_{miss}$ in the region $m^2_{miss}>0.85$ GeV$^2/c^4$ (left), where normalization events are abundant,  and to $O'_{NB}$ in the region $m^2_{miss}>0.85$ GeV$^2/c^4$ (right).  } \label{fit_belle_hadtag_leptau}
\end{figure*}

 In this measurement, hadronic tagging is employed where the $B_{\rm tag}$ is exclusively via one of 1149 hadronic modes in a hierarchal approach, and the $\tau$ decays leptonically, via $\tau^-\rightarrow \mu^-\bar{\nu}_{\mu}\nu_{\tau}$ or $\tau^-\rightarrow e^-\bar{\nu}_{e}\nu_{\tau}$. The normalisation mode in this case is $B\rightarrow D^{(*)}\ell\nu$ where $\ell=e,\mu$ and thus both the signal and normalization modes have the same final state.
The daughter of the $B_{\rm sig}$ is either a $D^0, D^+, D^{*0}$ or $D^{*+}$ and is combined with a daughter lepton $e$ or $\mu$.  $m^2_{miss}$ is then given by $m^2_{miss}=(p_{e^+e^-}-p_{\rm B_{tag}} -p_{D^*} - p_{\ell})^2$ and can be determined exactly.  $B\rightarrow D^{**}\ell\nu$ events are a challenging background with have the same signature as signal events in the high  $m^2_{miss}$ region. To suppress this background, a BDT is trained for each of the $D^{*+}\ell,D^{*0}\ell, D^0\ell$, and $D^+\ell$ samples where the main discriminating variable is $E_{ECL}$. $E_{ECL}$ is the sum energy of all neutral clusters in the event after the full signal selection is applied. It peaks at zero for signal and normalization events, as shown in Fig \ref{Eecl_belle}, whereas background events fill the entire spectrum.  After full reconstruction, the normalization yield is then extracted from a fit to $m^2_{miss}$ in the region $m^2_{miss}<0.85$ GeV$^2/c^4$ while the signal and background yields are extracted from $O'_{NB}$, the output of the trained BDT, in the $m^2_{miss}>0.85$ GeV$^2/c^4$  region. A large crossfeed between the $D$ and $D^*$ modes is present and is accounted for using MC simulation and added to the total signal and normalization yields, and is shown in Fig \ref{fit_belle_hadtag_leptau}. The final result is $R(D)=0.375\pm 0.064 \pm 0.026$ and $R(D^*)=0.293 \pm 0.038 \pm 0.015$. 
The leading systematic uncertainties arise from the modeling and composition of the $B\rightarrow D^{**}\ell\nu$ background. In terms of new physics extensions, the result is also consistent with the Two Higgs Doublet Model (2HDM) in the region $\tan\beta/m_{H^+} ~0.45\rm (c^2/GeV)$.

\subsection{$R(D^{(*)})$ with  hadronic tagging and hadronic $\tau$ decays \cite{belle_hadtag_hadtau_RDstar}}
At Belle, $R(D^{(*)})$ is also measured using hadronic tagging with $\tau^-\rightarrow\rho^-\nu_{\tau}$ and $\tau^-\rightarrow\pi^-\nu_{\tau}$. Here, the normalization mode is also chosen to be $B\rightarrow D^{(*)}\ell\nu_{\ell}$. Furthermore, a first measurement of the $\tau$ polarization is performed, which is an interesting observable with sensitivity to new physics. The $\tau$ polarization is defined by the following equation:
\begin{equation}
    P_{\tau}(D^*)=\frac{\Gamma^+(D^{(*)})-\Gamma^-(D^{(*)})}{\Gamma^+(D^{(*)})+\Gamma^-(D^{(*)})}
\end{equation}
where $\Gamma^+$ ($\Gamma^-$) is the $\bar{B}\rightarrow D^{*-}\tau^- \bar{\nu}_{\tau}$ decay rate with $\tau$ helicity of +1/2 (-1/2).  The $\tau$ polarization is determined as a function of $\cos\theta_{hel}$, the angle of the $\tau$ daughter meson momentum with respect to the direction opposite the $W$ momentum, in the rest frame of the $\tau$. With hadronic tagging, the $\tau\nu_{\tau}$ frame can be exactly determined, given the four momentum vector of the $B_{\rm sig}$.Signal selection consists of reconstructing charged or neutral $D$ or $D^*$ candidates. These are then combined with the $\tau$ daughter to form a $B_{sig}$ candidate. The signal region is then divided into two components: forward with $\cos\theta_{hel}>0$ and backward with $\cos\theta_{hel}<0$. The most significant irreducible background is from events with a misreconstructed $D^*$ candidate, denoted as fake $D^*$. After reconstruction, the signal and background yields are then extracted using a simultaneous fit to $E_{ECL}$ in 8 samples:  $B^+$ or $B^0$, $\pi$ or $\rho$, forward or backward. The resulting value is $R(D^*)=0.270\pm0.035 (\rm stat)^{(+0.028)}_{(-0.025)} (\rm sys)$. The leading systematics arise from the limited knowledge of the hadronci $B$ decays and the uncertainty in the shape and yield of the fake $D^*$ component. The $\tau$ polarization is also determined as $P_{\tau}(D^*)=2(N_{sig}^F-N_{sig}^B)/[\alpha(N_{sig}^F+N_{sig}^B)]$, where $N_{sig}^F$ ($N_{sig}^B$) is the number of events with positive (negative) $\cos\theta_{hel}$ and $\alpha$ is a constant dependent on the $\tau$ decay. The result  is $P_{\tau}(D^*)=-0.38\pm 0.51 (\rm stat)^{(+0.21)}_{(-0.16)} (\rm sys)$, and is the first measurement of the $\tau$ polarization. Both results agree with the SM expectation. 

\subsection{$R(D^{(*)})$ with  semileptonic tagging and leptonic $\tau$ decays}
\begin{figure*}[t]
\centering
\includegraphics[width=140mm]{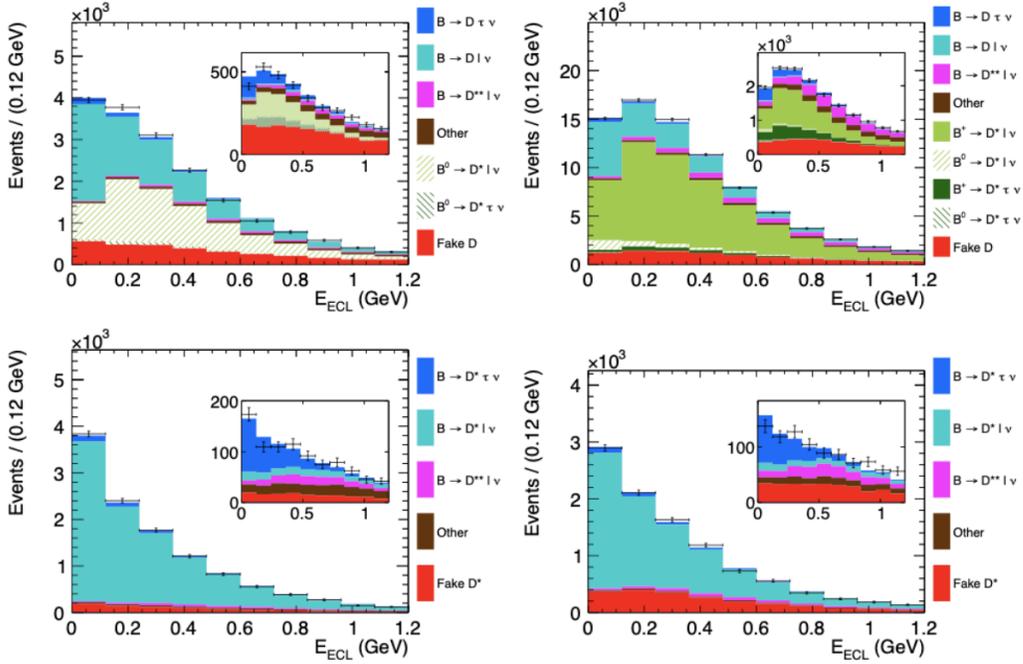}
\caption{Projections of the fit to $E_{ECL}$ for the $D^+\ell$ (top left), $D^0\ell$ (top right), $D^{(*+)}\ell$ (bottom left), and $D^{(*0)}\ell$ (bottom right) samples. The signal region, defined by the BDT output$>$0.9, is shown in the inset.  } \label{fit_eecl_sl_belle2}
\end{figure*}

The full Belle dataset is also used to measure $R(D^{(*)})$ with semileptonic tagging and leptonic $\tau$ decays. $B_{\rm tag}$ candidates are reconstructed via $B\rightarrow D^{(*)}\ell\nu$. On the signal side, a $D$ or $D^*$ candidate is reconstructed and combined with a lepton to form $B_{\rm sig}$. A cut of $E_{ECL}<1.2$ GeV is applied to suppress leading backgrounds from $B\rightarrow D^{**}\ell\nu$ events and fake $D^*$ components. A BDT is developed to distinguish  signal and normalization events from background events based on variables such as $m^2_{miss}$ and $E_{vis}$, the total visible energy in the event.
The  signal, normalization and background yields are then extracted from a two dimensional fit to $E_{ECL}$ and the BDT output, as shown in Fig. \ref{fit_eecl_sl_belle2}, in four samples: $D^{*+}\ell,D^+\ell,D^0\ell$ and $D^{*0}\ell$. The result is $R(D^*)=0.283\pm 0.018(\rm stat) \pm 0.014 (\rm sys)$ and $R(D)=0.307\pm 0.037(\rm stat)\pm 0.016(\rm sys)$. The leading uncertainties are from the limited size of the MC samples and the limited knowledge of the $B\rightarrow D^{**}\ell\nu$ background. This is the most precise measurement to date and is in agreement with the SM within 0.2$\sigma$ and 1.1$\sigma$ for $R(D)$ and $R(D^*)$ respectively. 

\section{ $R(D^{(*)})$ measurements at Belle II }
The Belle II experiment is the upgrade of its predecessor Belle with a projected luminosity that is 30$\times$ greater. Various improvements to the Belle II detector have been implemented to accommodate the increase in luminosity and maintain the high level of precision. Data taking at Belle II started in March 2019 and a total of 380 fb$^{-1}$ is already collected. Belle II will collect up to 480 fb$^{-1}$ before its first shut down, planned in summer 2022.  The short term plan is to confirm the current $R(D^{(*)})$ anomaly with the data collected before summer 2022 and present novel results on $R(X$), the corresponding inclusive ratio $R(X)=\frac{\mathcal{B}(B\rightarrow X\tau\nu_{\tau})}{\mathcal{B}(B\rightarrow X\ell\nu_{\ell})}$. To prepare for the Belle II measurements, a set of tools have been developed and will be discussed. 

\subsection{FEI algorithm}
As readily seen with the Belle measurements, $B$ tagging is an essential tool for $R(D^{(*)})$ measurements. At Belle II, a novel algorithm for hadronic and semileptonic $B$ tagging has been developed and is referred to as the Full Event Interpretation (FEI). It is a multivariate algorithm with a hierarchael approach, employing over 200 BDTs to reconstruct more than 10000 $B$ decay chains. The FEI is well developed within the Belle II software framework and has shown 30-50\%  improvement in efficiency compared to Full Reconstruction at Belle\cite{fei}.
\subsection{Particle Identification}
Global particle identification (PID)  is developed based on inputs from almost all the detector subsystems within the Belle II software framework. The PID performance is also evaluated using standard candle processes, such as $J/\psi \rightarrow e^+ e^-$ and $K_s^0 \rightarrow \pi^+ \pi^-$. At low momentum $p<1$ GeV/c, the mis-identification rate for leptons is usually larger. The low momentum range is of particular interest to semileptonic decays with $\tau$, since the $\tau$ decay products usually dominate that spectrum. To address this, an updated PID algorithm based on BDTs is developed, and utilizes information on the characteristics of the energy deposits of  low momentum leptons to improve the separation between leptons and hadrons. The updated PID algorithm reduces the misidentification rate for low momentum electrons by a factor of 10. 
\subsection{$E_{ECL}$ studies}
\begin{figure*}[t]
\centering
\includegraphics[width=135mm]{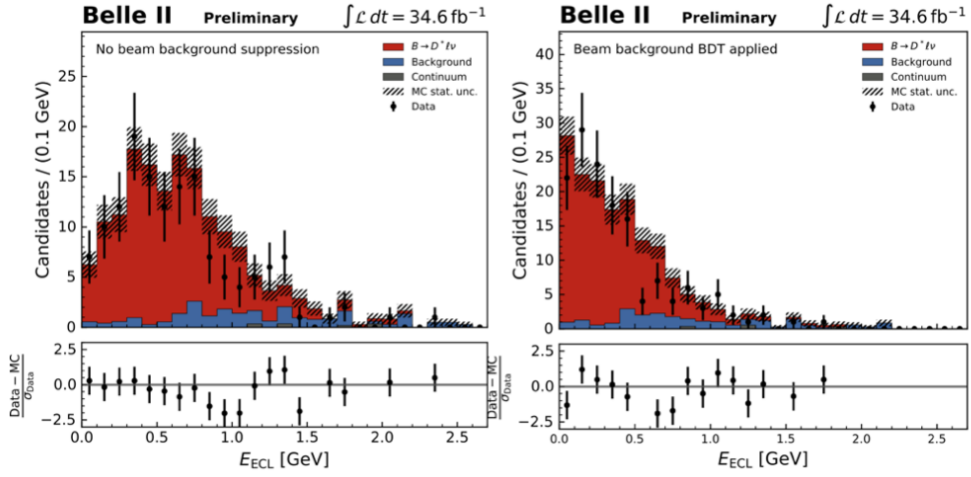}
\caption{The distribution of $E_{ECL}$ for simulated  signal and background, events before (left) and after (right) applying a cut on the BDT output for beam background suppression. The distribution of the Belle II data is shown as points with error bars.} \label{Eecl_belle2}
\end{figure*}

$E_{ECL}$ is a key variable for many semi-leptonic and missing energy measurements such as $R(D^{(*)})$. At Belle II, the higher luminosity achieved will increase the abundance of beam induced backgrounds, such as Touschek scattering. Further contributions to $E_{ECL}$ result from fake photons, which are  energy deposits or clusters resulting from hadronic showers and not attributed correctly to the mother hadron. These contributions push the $E_{ECL}$ distribution to non-zero values for properly reconstructed signal events, namely $B\rightarrow D^* \tau \nu_{\tau}$, and thus dilute the separation power of this variable.  To this effect, a BDT is developed to suppress contributions from beam backgrounds and fake photons in the $E_{ECL}$ distribution. This is achieved by employing cluster variables, such as  the cluster energy or polar angle, to distinguish between clusters related to real photons and clusters that originate from beam backgrounds or hadronic showers. The BDT is then tested on 34.6 fb$^{-1}$ sample of Belle II data, where $B^0\rightarrow D^{*+}\ell\nu_{\ell}$ events are selected. As shown in Fig \ref{Eecl_belle2}, the $E_{ECL}$ distribution does not peak at zero after the full selection. However, after applying a cut on the output of the BDT developed to suppress beam backgrounds, a peak at zero is recovered. This approach will be further developed to maintain and improve the separation power of $E_{ECL}$, a variable which exploits the exclusive control over the full event kinematics at $e^+e^-$ colliders. 

\subsection{Towards the future}
\begin{figure}[htb!]
\centering
\includegraphics[width=80mm]{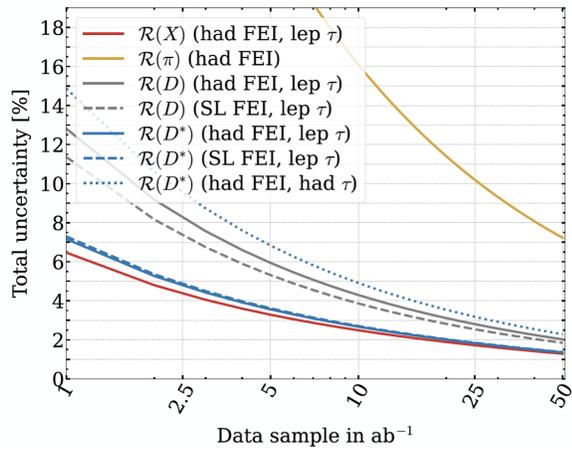}
\caption{Prospects for $R(D^{(*)}$ measurements with the projected luminosities at the Belle II experiment.} \label{snowmass}
\end{figure}

 Currently, at Belle II there are 3 measurements in progress:
\begin{itemize}
    \item $R(D^{(*)})$ with hadronic tagging and leptonic $\tau$ decays.
    \item $R(D^{(*)})$ with hadronic tagging and hadronic $\tau$ decays.
    \item $R(D^{(*)})$ with semi leptonic tagging and leptonic $\tau$ decays.
\end{itemize}
The short term goal is to confirm the current $B$ anomaly with ~0.5 ab$^{-1}$ of Belle II data. To check the robustness of the current data and the software framework developed, the branching fraction of $B^0\rightarrow D^{*+}\ell\nu_{\ell}$ is measured and is found to be consistent with the world average: $\mathcal{B}(B^0\rightarrow D^{*-}\ell^+\nu_{\ell})= 4.51 \pm 0.41 \rm(stat) \pm 0.27 \rm(sys) \pm 0.45_{\pi_s} $, where the last uncertainty is due to the efficiency of the low momentum pion in the decay $D^{*+}\rightarrow D^0 \pi^+$. \\
The first results for $R(D^{(*)})$ are expected in the upcoming few months. As the size of the Belle II data sample increases, the sensitivity of the  $R(D^{(*)})$ measurements will improve as shown in Fig \ref{snowmass} and will reach a total uncertainty of 2\% at 50 ab$^{-1}$. The Belle II data set  can provide the answer to whether or not new physics contributions are present in $R(D^{(*)})$, as well as various other avenues for the new physics searches. 

\begin{acknowledgments}
This document is adapted from the ``Instruction for producing FPCP2003
proceedings'' by P.~Perret and from eConf templates~\cite{templates-ref}.
\end{acknowledgments}

\bigskip 

\end{document}